# What is in a Scent? Understanding the role of scent marking in social dynamics and territoriality of free-ranging dogs

Sourabh Biswas[1], Kalyan Ghosh[1], Swarnali Ghosh[2], Akash Biswas[2] & Anindita Bhadra[1*]

[1]Behavior and Ecology Lab, Department of Biological Sciences, Indian Institute of Science Education and Research Kolkata, Mohanpur, West Bengal, India.

[2]Department of Microbiology, Kalyani Mahavidyalaya, Kalyani, West Bengal, India.

sb18rs107@iiserkol.ac.in, kg23rs078@iiserkol.ac.in, swarnalighosh3554@gmail.com, akashbiswaslm12@gmail.com, abhadra@iiserkol.ac.in*

*Corresponding author address:

Behavior and Ecology Lab, Department of Biological Sciences, Indian Institute of Science Education and Research Kolkata, Mohanpur, West Bengal, India.

**ORCID:**

Sourabh Biswas: https://orcid.org/0000-0001-6187-8106

Kalyan Ghosh: https://orcid.org/0000-0002-2602-8912

Swarnali Ghosh: https://orcid.org/0009-0005-1715-3411

Akash Biswas: https://orcid.org/0009-0006-6788-4690

Anindita Bhadra: https://orcid.org/0000-0002-3717-9732

**Abstract:**

Scent marks play a crucial role in both territorial and sexual communication in many species. We investigated how free-ranging dogs respond to scent marks from individuals of different identities in terms of sex and group, across varying strategic locations within their territory. Both male and female dogs showed heightened interest in scent marks compared to control, exhibiting stronger territorial responses,. with males being more territorial than females. Overmarking behaviour was predominantly observed in males, particularly in response to male scent marks and those from neighbouring groups. Behavioural cluster analysis revealed distinct responses to different scent marks, with neighbouring group male scents eliciting the most distinct reactions. Our findings highlight the multifaceted role of scent marks in free-ranging dog communication, mediating both territorial defence and intrasexual competition. The differential responses based on the identity and gender of the scent-marker emphasize the complexity of olfactory signalling in this species. This study contributes to understanding the social behaviour of dogs in their natural habitat, and opens up possibilities for future explorations in the role of olfactory cues in the social dynamics of the species.

**Introduction:**

Chemical communication, particularly through scent marking via urine, faeces, and glandular secretions, is a widespread phenomenon in the biological world, employed by a diverse range of organisms including plants, animals, and even microorganisms (Ferkin et al., 2017; Hurst et al., 2001). These scent marks are hypothesized to act as honest signals (Zahavi, 1975), which, by definition, must be costly or difficult to produce and maintain, ensuring that they are reliable. In the context of chemical communication, honest signals conveyed through scent marking could potentially encode information about an individual's species, age, sex, reproductive status, health condition, or even unique identity (Brown & Macdonald, 1985; Grose, 2011; Halpin, 1980). Following on the lines of the honest signal hypothesis, individuals with better reproductive health or other physiological markers of health are expected to maintain potent and persistent scent marks. This mechanism ensures that scent marks truthfully communicate territorial ownership, reproductive status, and individual identity (Hurst & Rich, 1999; Müller-Schwarze & Silverstein, 2013; Soso et al., 2014).

Among the information conveyed by scent marks, individual identity is particularly unique, requiring scents of sufficient complexity to enable sufficient variation between individuals. Many mammalian species have demonstrated the ability to discriminate individual scent marks. In group-living species, this ability plays a crucial role in social interactions and maintaining group cohesion. For instance, ring-tailed lemurs (*Lemur catta*) (Mertl, 1975; Palagi & Dapporto, 2006) and African dwarf mongooses (*Helogale undulata rufula*) (Rasa, 1973) rely on individual scent marks for intra-group recognition and communication of social status. Similarly, wolves (*Canis lupus lupus*) and dogs (*Canis lupus familiaris*) (Johnston,

2008) use urine, feces, and glandular secretions to mark territories and signal individual identity within packs.

Solitary living species also use individual scent marks for critical communication, particularly in the context of mating and territory marking. Examples include banner-tailed kangaroo rats (*Dipodomys spectabilis*) (Randall, 1987), Mongolian gerbils (*Meriones unguiculatus*) (Halpin, 1974), golden hamsters (*Mesocricetus auratus*) (Johnston et al., 1993), small Indian mongooses (*Herpestes auropunctatus*) (Gorman, 1976) and tigers (*Panthera tigris*) (Smith et al., 1989), that use individual scent marks for territory marking and mating-related signalling.

Scent marking plays a crucial role in reducing and resolving conflicts in many species. It benefits both the marker and the investigator by allowing the identification of high-threat, high-status, or extremely fit competitors, thereby avoiding direct confrontations. Scent marking serves as a means of territorial defence, social signalling, and conveying information about dominance and threat (Arakawa et al., 2008; Cafazzo et al., 2012). Studies have shown that scent marking can be influenced by social experience, context specificity, and handler beliefs (Lit et al., 2011; Roullet et al., 2011). Furthermore, scent marking behaviours are linked to social recognition, familiarity, and responses to social signals (Arakawa et al., 2007, 2008).

Canids with diverse social structures engage in countermarking using urine and feces, both within and between sexes, social groups, and occasionally even across species. This behaviour is observed in species such as the domestic dog (*Canis lupus familiaris*), African wild dog (*Lycaon pictus*), swift fox (*Vulpes velox*), crab-eating fox (*Cerdocyon thous*), gray wolf (*Canis lupus lupus*), golden jackal (*Canis aureus*), coyote (*Canis latrans*), bat-eared fox (*Otocyon megalotis*), and bush dog (*Speothos venaticus*) (Bowen & Cowan, 1980; Brown & Macdonald, 1985; Darden et al., 2008; Rothman & Mech, 1979; Ryon & Brown, 1990). The primary contexts for canid countermarking are intersexual interactions during courtship and between members of an established pair (Brown & Macdonald, 1985; Porton, 1983; Rothman & Mech, 1979). This behaviour is believed to facilitate courtship, the formation and maintenance of pair bonds, mate guarding, and territory marking (Asa et al., 1990; Bowen & Cowan, 1980; Brown & Macdonald, 1985; Dunbar & Buehler, 1980; Rothman & Mech, 1979), although these proposed functions have yet to be empirically tested.

Most studies on scent marking in dogs, particularly through raised-leg urination, focus on the relationship between mating, gonadal hormones, and urinary behaviours (Cafazzo et al., 2012). Urination rates typically increase in both sexes during the oestrous season (Pal, 2003) and with the administration of exogenous gonadal hormones (Beach, 1974; I. F. Dunbar, 1978). Within established groups, urine marking and investigation are crucial for mating, with males generally investigating and responding to urine marks more frequently than (Doty & Dunbar, 1974; Dunbar & Buehler, 1980; Dunbar, 1977, 1978; Ranson & Beach, 1985). However, responses to urine from unfamiliar dogs can vary. While aggression and competition can occur among group members, established social relationships, hierarchies, and appeasement behaviours often mitigate intragroup aggression (Beach et al., 1982; Schenkel, 1967; Scott & Fuller, 2012).

The response to unfamiliar scent marks is vital in territorial behaviour, with animals showing increased vigilance and aggression to defend territories and signal intergroup threats (Christensen et al., 2016; Green et al., 2021). This reflects the "dear enemy" effect, where aggression is reduced towards familiar neighbours but heightened towards intruders (Le Roux et al., 2008). Increased marking in response to rival scents further underscores scent's role in social interactions and territorial defence (Massen et al., 2016; Oliveira & Macedo, 2010), helping maintain social hierarchies and boundaries.

Free-ranging dogs in India occupy varied habitats, living in close proximity with humans, in their own social groups (Sen Majumder et al., 2014, 2016). The social groups are often comprised of family members, and are highly territorial. They mate promiscuously and show extensive maternal care and alloparenting (Paul & Bhadra, 2018). Even while resting, they choose strategic sites within their territories, that can allow them to guard resources and remain vigilant, if necessary (Biswas et al., 2023). Hence, social interactions and cohesion are important factors in the daily lives of free-ranging dogs. They are known to show urine marking, both by raising their legs and squatting (Pal, 2003). Living in the human-dominated urban landscape exposes these dogs to myriad challenges, one of which is navigating through a complex olfactory environment. Under these circumstances, it is interesting to understand the specific role of urine marking, the context in which it is performed and what signals it might convey to the dogs.

We carried out a field based, population level experiment to investigate whether free-ranging dogs can distinguish between the scent marks of their own group and neighbouring groups, and to understand how these scent marks might trigger territorial responses in dogs across different strategic locations within their territory. Additionally, we aimed to explore the countermarking behaviour of dogs, if any, in response to scent marks from the same group or neighbouring groups, and to explore the role of gender in these responses. We hypothesized that dogs would spend more time investigating urine from unfamiliar individuals compared to urine from their groupmates. Additionally, we expected that males would exhibit a stronger interest in female urine due to sexual attraction.

**Methodology:**

*Study sites and animals:*

The experiments were conducted on groups of free-ranging dogs across 25 different locations (Fig.S1 & Table S1) in the Nadia and Bardhaman districts of West Bengal, India, from

March 2022 to November 2023. All experiments were conducted under daylight, between 0700 to 1800 hours.

Prior to commencing of the experiments, extensive censuses and surveys were conducted to comprehensively assess the demographics and distribution of dog groups within the study area. This involved meticulous observations of group dynamics and the identification of key locations such as resting sites, resource areas, and territorial boundaries for each focal group.

*Preparation and experimental setup:*

Scent marks were collected from both male and female dogs onto 15 cm by 15 cm white cotton cloth pieces. Commercially available new white cloth was prepared for the experiment by soaking in water for 24 hours, washing with tap water, and drying in sunlight before being cut to size.

To facilitate the controlled presentation of scent marks for observation and analysis, we designed a stainless-steel stand. This stand measured 50 cm in height and had a 50 cm extended arm with a hook for attachment (Fig. S2). During the experimental trials, dog groups were presented with either a test cloth with dog scent mark or a control cloth, with a splash of water.

*Collection of scent mark:*

Upon reaching the study site, experimenters patiently awaited natural scent marking behaviours within dog territories. Scent marks were then carefully collected by using the cotton cloth to swab the marked surfaces, ensuring minimal contamination. The collected scent marks were stored in airtight aluminium foil and used for experiments within 45 minutes of collection. The experimenters always used clean gloves for the collection and presentation of scent marks.

*Protocol:*

The experiment comprised of presenting scent marks to dog groups and were broadly divided into two categories: intragroup and intergroup scent marks. In the former, scent marks collected from individual dogs were reintroduced within their own group's territory, while in the latter, the scent marks were presented to neighbouring dog groups. Further, for each category of experiments, there were three sub-categories, based on the location within the territory where the scent mark was presented. Based on earlier surveys, the territories were clearly mapped and the scent marks were presented (i) near resources, (ii) at resting sites, and at the territorial boundaries.

*Experimental procedure:*

The experimental procedure entailed positioning the setup in front of a dog group and audibly calling out "aye aye aye" (Bhattacharjee et al., 2017) for three seconds to capture the dogs' attention, before moving away. Video recording commenced simultaneously with the setup introduction, documenting the entire interaction. Each experiment had a duration of three minutes from the call, during which all behaviours, including the presence of nearby dogs, were recorded.

*Data decoding:*

Behavioural responses of individual dogs interacting with the setup were decoded from the videos by a single experimenter. A dog was defined as a responder only if it sniffed the setup (either the stand or cloth). The behaviours of all dogs within three dog body-lengths of the setup were recorded for the duration of the experiment. All the behavioirs were categorized based on the ethogram in Banerjee & Bhadra (2022).

**Statistical Analysis**

All statistical analyses were conducted in R Studio (version 4.2.0). The Wilcoxon rank-sum test was used to compare latency between the test and control groups, as well as among genders of the responders for both the test and control trials. The Kruskal-Wallis test was used to compare latency across different positions (resource areas, resting areas, and territorial boundaries) for performing the experiments.

To evaluate the effects of sample category (scent mark/control), experimental group (same group/neighbouring group), and experimental positions on the duration of investigation by the first responder and the average investigation time of all responders, a Generalized Linear Model (GLM) with a Gamma distribution was performed using the 'lme4' package in R (Bates et al., 2015).

The chi-square test was used for comparing the proportions of male responses to male scent marks versus female responses to male scent marks, and for male responses to neighbouring group scent marks versus same group scent marks. Fisher's Exact Test compared proportions between male and female responses to female scent marks, and between male responses to male and female scent marks.

Additionally, GLMs were employed to investigate the effects of sample categories, experimental groups, and experimental positions on territorial Behaviour response scores. Response scores were based on the magnitude of territorial behaviour observed in the experimental setup. In each experiment, each responding dog received a territorial response score, which was determined by the total investigation time and other territorial behaviours (Table 1) exhibited during that experiment. The scores for all behaviours shown by a particular dog were then summed up to obtain the final score.

The null and full models were compared for all analyses. Model assumptions, dispersion, and residual diagnostics were checked using the "performance" package (Lüdecke et al., 2021) in R. All models met the assumptions, and the residuals were satisfactory. The alpha level was set at 0.05 throughout the analysis.

| Behaviour category | Behaviour description | Score |
|---|---|---|
| Investigation | Sniffing < 5 sec | 1 |
| Investigation | Sniffing 6-15 sec | 2 |
| Investigation | Sniffing 16-30 sec | 3 |
| Investigation | Sniffing > 30 sec | 4 |
| Vigilance | Alert | 5 |
| Vocalisation | Aggressive barking | 6 |
| Scent marking | Scent marking on areas other than setup | 7 |
| Overmarking | Overmarking on experimental setup | 8 |
| Ground scratching | Ground scratching followed by overmarking | 9 |

*Table 1: Scoring criteria for territorial behaviours shown by dogs. This table details the categories of behaviours observed during the experiment, along with their descriptions and the corresponding scores assigned. These scores reflect the intensity of the territorial response, which were summed up to calculate the total territorial response score for each dog in the experiment.*

The Kruskal-Wallis rank sum test, followed by Dunn's test, was utilized to compare the differences in the proportions of time invested across the different categories of territorial behaviour shown by the responders.

A behavioural cluster analysis was conducted to understand the relationship between the behavioural profiles of nine unique responder categories. These categories were based on responders' gender (male/female) and their responses to experimental cues (scent mark and control) across various experimental categories (neighbouring/same group) and sample genders (male/female). For instance, categories included male responders toward neighbouring male scent marks (M_Neigh_M), male responders toward control (M_control), and similar variations.

Hierarchical cluster analysis was performed using the 'vegan' package (Oksanen, 2010) in R. Initially, a dissimilarity matrix was created between responder category based on their behaviour frequency, utilizing the Bray-Curtis method. This dissimilarity matrix was then employed to construct two cluster dendrograms: one including all behaviours and another focusing solely on territorial behavioural responses.

**Results:**

**Latency**

We conducted a total of 302 trials. Out of these, 192 trials elicited a response by dogs. The latency of the first responder towards the set-up was comparable across categories. Specifically, there were no significant differences observed between the test and control groups (Wilcoxon rank sum test: $W = 3249$, $p = 0.3114$), genders ($W = 897$, $p = 0.2091$), group categories (same vs. neighbouring: $W = 2823.5$, $p = 0.325$), or experimental positions (resource, resting site, territorial boundary: Chi-square test: $\chi^2 = 3.7887$, $df = 2$, $p = 0.1504$).

**Investigation time of the first responder**

GLM revealed that the investigation time of first responder varied significantly depending on the sample category (scent mark/control) ($t = 4.157$, $p < 0.001$). Individuals exhibited a longer investigation time for scent marks compared to control (Fig. 1a & Table S2a). The experimental group (same group/neighbouring group) had a significant impact on investigation time ($t = -3.593$, $p < 0.001$). Investigation time was shorter for scent marks from the own groups of the responders as compared to those from neighbouring groups (Fig. 1b & Table S2b). Additionally, individuals showed no difference ($p > 0.05$) in investigation time for responses to scent marks placed in different positions in the territory.

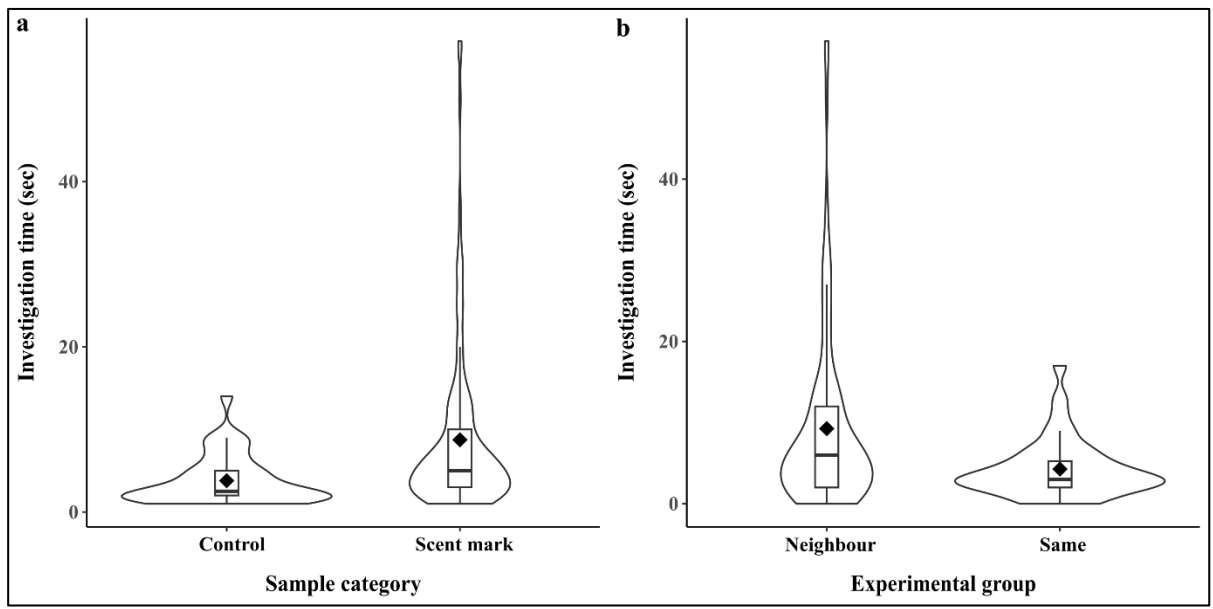

*Fig. 1: Comparison of investigation time of first responder.* *This box-and-whisker plot, embedded within a violin plot, illustrates the differences in investigation time among first responder dogs across categories. a) Comparison of investigation time across sample categories; b) Comparison of investigation time across experimental groups. The black line represents the median, the black diamond indicates the mean, the rectangle highlights the interquartile range (25th and 75th percentiles), and the whiskers show the range of the data. The shape of the violin reflects the distribution of the data.*

**Investigation time overall**

Generalized linear models (GLMs) were fitted to investigate the factors influencing the average investigation duration of all responders in the studied population. The results for the population were similar to those shown by the first responders. The duration of investigation varied significantly depending on the sample category (scent mark/control) (t = 5.102, p < 0.001). Individuals exhibited a longer duration of investigation for scent marks compared to control (Fig. 2a & Table S3a). The experimental group (same group/neighbouring group) had a significant impact on the duration of investigation (t = 3.591, p < 0.001). Investigation time was higher for neighbouring group scent mark as compared to own group (Fig. 2b & Table S3b). However, there was no difference (p > 0.05) in investigation time across different experimental positions in the territory.

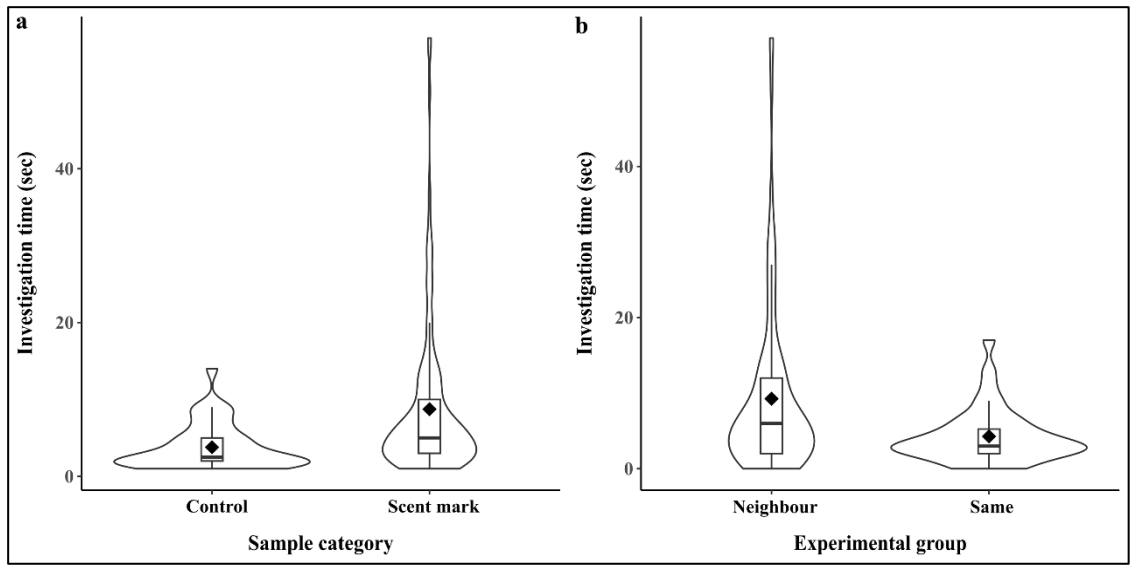

*Fig. 2: Comparison of average investigation time of the responders. This box-and-whisker plot, embedded within a violin plot, illustrates the differences in average investigation time among responder dogs across categories. a) Comparison of average investigation time across sample categories; b) Comparison of average investigation time across experimental groups. The black line represents the median, the black diamond indicates the mean, the rectangle highlights the interquartile range (25th and 75th percentiles), and the whiskers show the range of the data. The shape of the violin reflects the distribution of the data.*

**Over marking**

Among the 192 events where dogs responded to our experiment, dogs were observed to overmark 37 times. Of these 37 instances, 29 were directed at the test sample, and 8 were directed at the control (Fig. 3). Fisher's exact test revealed no significant difference in proportions between male and female responses to female scent marks (p > 0.05) and between male responses to male and female scent marks (p > 0.05).

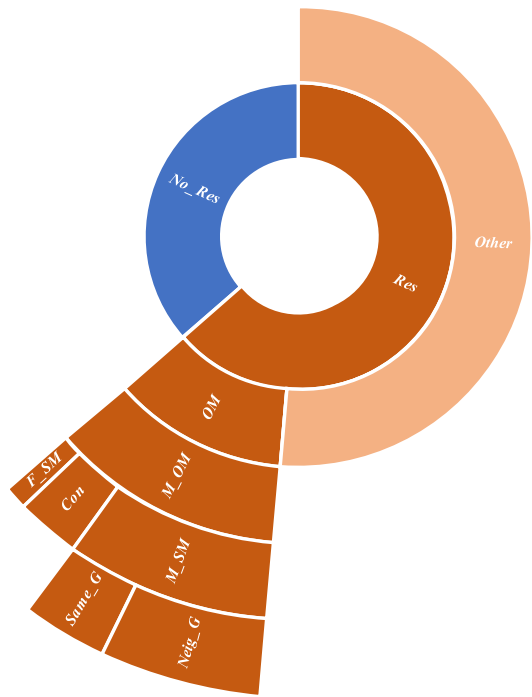

*Figure 3: Overmarking in dogs: Comparison of dogs' behavioral responses (Res) to non-responses (Non_Res) in the experimental setup, illustrated by a sunburst plot. The plot first compares the proportion of overmarking response (OM) with other responses by the dogs. It then further breaks down the proportion of overmarking by males (M_OM) in response to different stimuli: control (Con), male scent marks (M_SM), and female scent marks (F_SM). Additionally, the proportion of male overmarking responses to male scent marks (M_SM) is compared between responses toward same group scent marks (Same_G) and neighboring group scent marks (Neig_G).*

The chi-square test indicated a significant difference in proportions between male responses to male scent marks and female responses to male scent marks ($\chi^2 = 40.761$, df = 1, $p < 0.001$), with males (0.38) responding more frequently than females (0.00). The chi-square test also showed a significant difference in proportions between male responses to neighbouring group scent marks and same group scent marks ($\chi^2$ = 3.968, df = 1, p < 0.05), with males responding to neighbouring groups (0.64) more frequently than same group (0.31).

**Behavioural response score**

Generalized linear models (GLMs) were fitted to investigate the factors influencing territorial behavioural response scores. The GLM analysis revealed a significant effect of sample category (scent mark/control) on the behavioural response score. Individuals exhibited higher territorial behaviour responses in the test condition compared to the control (t = 2.812, p < 0.001) (Fig. 4a & Table S4a).

Additionally, the GLM showed no significant difference (p > 0.05) in territorial responses between same-group and neighbouring-group scent marks. There was no difference (p > 0.05) in territorial response across different experimental positions in the territory (ESM). Furthermore, males showed higher territorial responses compared to females (t = 3.53, p < 0.001) (Fig. 4b & Table S4b).

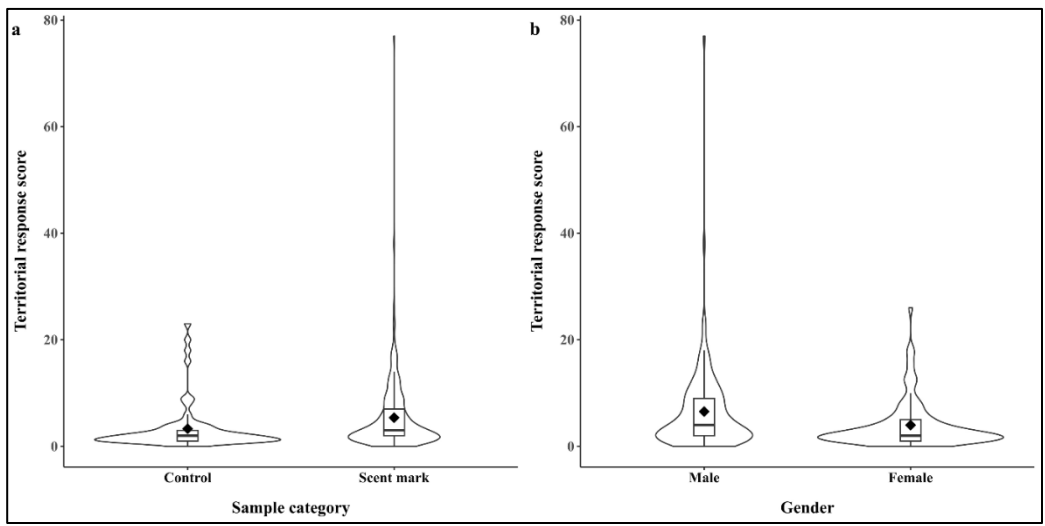

*Figure 4: Comparison of territorial response score of the responders. This box-and-whiskers plot embedded within a violin plot illustrates the differences in territorial response score among responder dogs across categories. a) Comparison of territorial response scores across sample categories; b) Comparison of territorial response scores across responder gender.*

**Behavioural Cluster**

The Kruskal-Wallis rank sum test revealed a significant difference in the proportions of time invested across behavioural categories ($\chi^2 = 221.51$, df = 3, $p < 0.001$). Post-hoc pairwise comparisons using Dunn's test indicated that responders spent significantly more time in investigation ($p < 0.001$) than in aggression, overmarking, and vigilance behaviours.

In the behavioural cluster analysis (Fig. 5), we identified three distinct clusters at a height between 0.8 and 1. Males showed distinct behaviours toward same-group female scent marks.

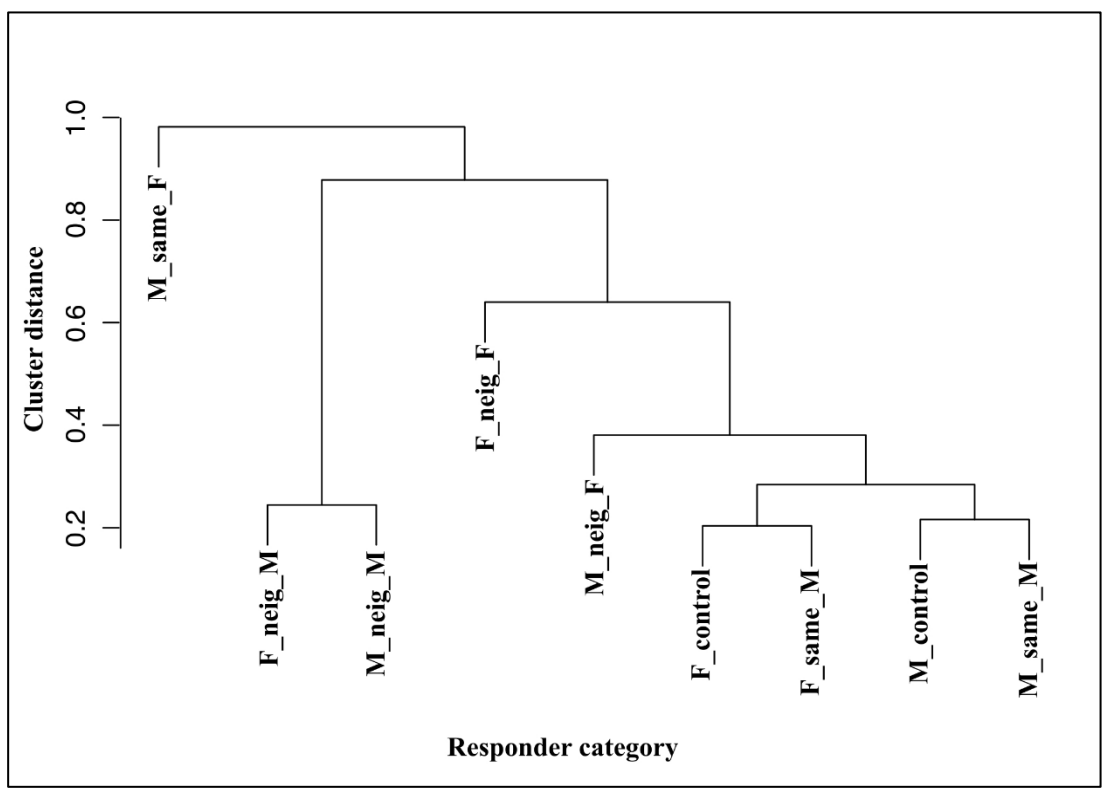

*Figure 5: Behavioural clustering of dogs exposed to different olfactory cues.*
*This hierarchical clustering dendrogram illustrates the grouping of behavioural responses exhibited by dogs when exposed to various olfactory cues: a control scent (water), scent marks from their own group (same), and scent marks from neighbouring groups (neig). In the labels, 'F' denotes female and 'M' denotes male individuals. For example, **'M_same_F'** represents the response of a male dog to the scent mark of a female dog from its own group.*

Both males and females exhibited similar behaviours toward neighbouring male scent marks. Other categories showed distinct clusters. Males' responses to same-group male scent marks and control conditions were behaviourally similar. Similarly, females' responses to same-group male scent marks and control conditions were also similar. In contrast, females' responses to neighbouring female scent marks and males' responses to neighbouring female scent marks displayed different behavioural patterns.

**Discussion**

Scent-marking behaviour in domestic dogs is a well-documented form of communication that plays a crucial role in territorial signalling and social interactions (Cafazzo et al., 2012). Dogs use scent marks to convey information about their presence, status, and territorial boundaries to conspecifics. The investigation of scent marks by dogs can be linked to their social structure, competition for resources or mates and the maintenance of territorial integrity within their groups. This behaviour aligns with the territorial defence hypothesis, where scent marking serves as a means of demarcating territorial boundaries and deterring potential intruders (Cafazzo et al., 2012).

Our study revealed interesting differences between the responses of free-ranging adult dogs to scent marks from individuals of different identities. As expected, the test conditions on the whole induced more responses than the control, suggesting that the responses of the dogs were indeed to the scent mark, rather that the novel set-up. This is important, as dogs are generally known to show neophilia, and in the context of scavenging, free-ranging dogs have been shown to display neophilia (Bhattacharjee et al., 2024). Both male and female dogs investigated scent marks longer than control marks, suggesting that the response to scent marks is not restricted to any sex. Scent marks are likely to be most relevant to both sexes in the context of territoriality, as the dogs spent more time investigating scent marks from neighbouring groups compared to their own group.

The dogs prioritize investigating scent marks over aggression, overmarking, and vigilance, emphasizing the importance of olfactory information in their social and territorial interactions. Investigating scent marks allows dogs to gather crucial details about identity, reproductive status, health, and social rank (Bekoff, 2001; Johnston, 2008), helping them make informed decisions and avoid conflicts (Gosling & Roberts, 2001). The lower focus on aggression, overmarking, and vigilance suggests that dogs use investigation to assess social and environmental contexts efficiently. Since aggression and overmarking are energy-intensive and vigilance is "mentally taxing" (Gosling & Roberts, 2001), prioritizing investigation helps dogs optimize their responses.

Contrary to earlier findings of a strong sexual dichotomy (Dunbar, 1978; Ranson & Beach, 1985), Lisberg & Snowdon (2009) found that both male and female dogs investigated urine equally, with females showing more interest in unfamiliar urine. Our findings align well with this, as both sexes investigated scent marks equally in our study. Lisberg & Snowdon (2011) reported that overmarking is primarily done by males, particularly high-status males, and occurs in response to both familiar and unfamiliar urine. Adjacent marking, on the other hand, is done equally by both sexes, isn't linked to social status, and only happens in response to unfamiliar urine. In this study too, overmarking behaviour was predominantly observed in males. Both genders overmarked female scent marks at similar rates, but males were more likely to overmark neighbouring group scent marks compared to their own group. The higher tendency for males to overmark compared to females can be explained both in the context of mating and territoriality. Overmarking allows individuals to assert dominance and establish territorial boundaries, which are crucial for mating and resource defence, and can have underlying roles in mating success of the males. This behaviour is consistent with observations in other canid species, such as wolves and coyotes (Asa et al., 1985; Bowen & Cowan, 1980).

In our study, both sexes overmarked female scent marks equally, suggesting that these marks may carry significant social information related to reproductive status and/or identity. Males, however, did not differentiate between male and female scent marks, possibly indicating that male overmarking also serves to maintain social cohesion and group identity (Allen et al., 2015). Further experiments coupled with long-term observations are needed to fully understand this behaviour.

The free-ranging dogs exhibited stronger territorial responses to scent marks than to controls, as expected, with similar responses for scent marks from neighbouring groups and their own group. Males displayed higher territorial responses than females. Investigation of the scent mark was the most frequent behavioural response observed. Behavioural cluster analysis revealed that the responses to the different scent marks presented were determined by the individual identities. Neighbouring group males elicited similar responses from both males and females, and these responses were the most distinct from those shown to the control and same group males by both sexes. The neighbouring group female scent marks elicited distinctly different responses from both males and females as compared to neighbouring as well as same group males. This strongly shows a distinct pattern in the behavioural responses to the different cues presented, and confirms that scent marks are likely to play dual roles in sexual as well as territorial communication.

Males overmarked more frequently on the scent marks of neighbouring groups than on those of their own group. This supports the hypothesis that overmarking is a strategy for intergroup competition and territory defence. By overmarking rivals' scent marks, males may disrupt their olfactory cues, confuse their spatial orientation, and weaken their territorial claims, similar to strategies observed in other territorial mammals (Hurst & Rich, 1999). The differential overmarking of intra-group versus inter-group scent marks likely reflects a balance between internal cooperation and external competition. In cohesive social structures like the free-ranging dog populations, maintaining clear territory boundaries is essential for group stability and resource access. Overmarking serves both as a deterrent to rival groups and a signal of group unity (Müller & Manser, 2008).

Our study highlights significant differences in how males and females respond to scent marks, underscoring the complexity of their social and territorial behaviours. Males exhibited distinct behaviours towards same-group female scent marks, suggesting these marks carry crucial social and reproductive information. Notably, males engaged in more frequent overmarking, which is likely driven by both territorial defence and sexual status signalling (Gese & Ruff, 1997). In contrast, females primarily showed territorial responses, with less emphasis on sexual status signalling. Both sexes responded similarly to neighbouring male scent marks, indicating a shared focus on addressing potential territorial threats. Similar responses to same-group male scent marks and control conditions suggest these marks are not seen as immediate threats, possibly due to social tolerance within the group (Gese & Ruff, 1997). Distinct behavioural patterns were observed for neighbouring female scent marks, with males and females showing different responses. This variation could reflect intra-sexual competition or social hierarchy (Mitchell, 2017). Our findings underscore that scent marking serves as a complex form of olfactory communication, essential for maintaining group cohesion, defining territorial boundaries, and managing intra- and inter-group relationships (Becker et al., 2012; Hurst & Rich, 1999).

This, to the best of our knowledge, is the first experimental study to understand the responses to specific scent marks in free-ranging dogs. This study has provided insights into the complex repertoire of responses to scent marks, shown by adult free-ranging dogs, in the overlapping contexts of territoriality and reproduction. Often, lab-based studies are aimed to understand behaviours in very specific context, but in real life, animals face multiple cues and undergo decision-making processes in complex scenarios. This study aimed to replicate such scenarios, and the diverse responses were found to be specific to sex and group identity, suggesting a strong role of scent marks in social cohesion and territoriality, in addition to their importance in reproduction. Future research should investigate the specific chemical components of scent marks that trigger distinct behaviours and their impact on social dynamics and territoriality over time, and in different life history stages of the free-ranging dogs.

**Acknowledgements:**

The authors extend their gratitude to Ms. Hindolii Gope, Ms. Srijaya Nandi and Dr. Dipankar Debnath for their valuable assistance in the field experiments. They also acknowledge the Indian Institute of Science Education and Research Kolkata for providing the necessary infrastructural support.

**Funding:**

SB would like to thank the University Grants Commission, India for providing him doctoral fellowship. The project was partially funded by the Janaki Ammal Award grant BT/HRD/NBA-NWB/39/2020-21 (YC-1), to AB by the Department of Biotechnology, India.

**Author contributions**

SB, KG, SG and AB carried out the field work. SB and KG curated the data, SB carried out the data analysis and wrote the manuscript. AB conceived the idea, supervised the work, acquired funding, reviewed and edited the manuscript.

**Data availability statement**

Data supporting the results will be archived.

**Conflict of Interest Information**

Authors have no conflict of interest.

**Supplementary Materials:**

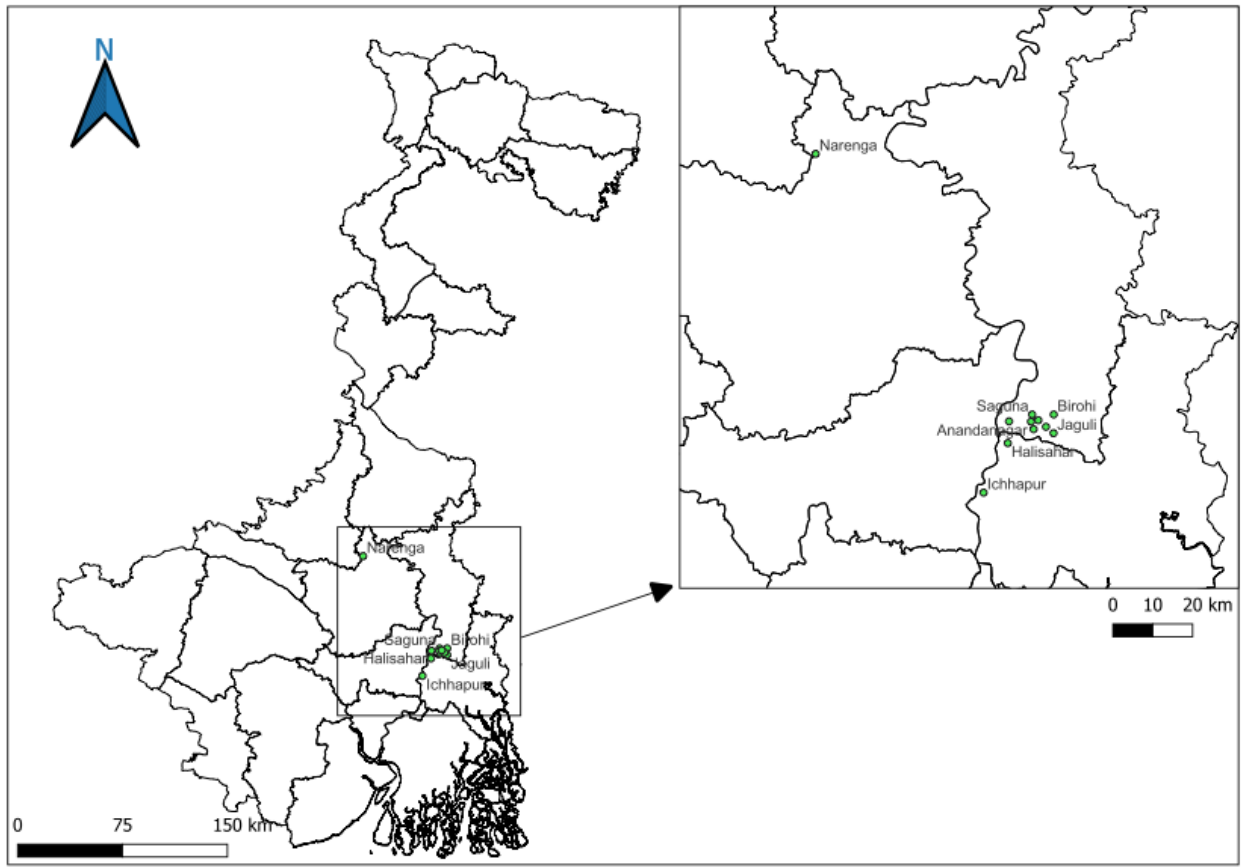

**Fig. S1: Map of study sites:**

A map of West Bengal, India, highlighting the 26 study sites across different districts. In Nadia district, we covered 7 locations in Kalyani, 7 in Gayeshpur, 3 in Saguna, 2 in Birohi, and one location each in Anandanagar, Raghunathpur, Jaguli, and IISER Kolkata campus. In North 24 Parganas district, we covered one location each in Halisahar and Ichhapur. Additionally, one location was covered in Narenga, located in Purba Bardhaman district.

**Table S1: Description of Study Sites:**

This table provides a detailed description of the study sites, including GPS locations, habitat types, and specific habitat features. All sites are located in West Bengal, within the Indo-Gangetic plain, which experiences similar climatic conditions. The region is characterized by a tropical wet and dry climate, with annual rainfall occurring primarily between mid-July and mid-September.

| Location | Latitude | Longitude | Habitat type | Description |
|---|---|---|---|---|
| Kalyani | 22.975084 | 88.434509 | Urban | Kalyani is a planned urban city in Nadia district, West Bengal. It is also a municipality. |
| IISER Kolkata | 22.964890 | 88.526614 | Semi-urban | Academic and research institute situated at Mohanpur, Nadia, in a semi-urban municipal area. |
| Gayeshpur | 22.958457 | 88.495431 | Semi-urban | Gayeshpur is a town and municipality in Nadia district, West Bengal. It is part of the Kolkata Metropolitan Development Authority, with well-planned streets. |
| Saguna | 22.991169 | 88.491441 | Semi-urban | Saguna is a small town with a well-planned network of streets. |
| Birohi | 22.994129 | 88.544629 | Rural | Birohi is a village located in Haringhata subdivision of Nadia district, West Bengal. |
| Anandanagar | 22.976282 | 88.485437 | Semi-urban | Anandanagar is a small town with a well-planned network of streets. |
| Raghunathpur | 22.975520 | 88.496566 | Semi-urban | Raghunathpur is a small town with a well-planned network of streets. |
| Jaguli | 22.946662 | 88.548111 | Semi-urban | Jaguli is a village located in Haringhata subdivision of Nadia district, West Bengal. It is a small town with some agricultural fields, within Haringhata Municipality. |
| Halisahar | 22.926068 | 88.424362 | Urban | Halisahar is a city and municipality in North 24 Parganas district, West Bengal. It is part of the Kolkata Metropolitan Development Authority. |
| Ichhapur | 22.800193 | 88.375034 | Urban | Ichhapur is a locality in North Barrackpur Municipality, North 24 Parganas district, West Bengal, and is part of the Kolkata Metropolitan Development Authority. |
| Narenga | 23.252451 | 87.839693 | Semi-urban | Narenga is a small village located on the banks of the Ajay River in Purba Bardhaman District, West Bengal. |

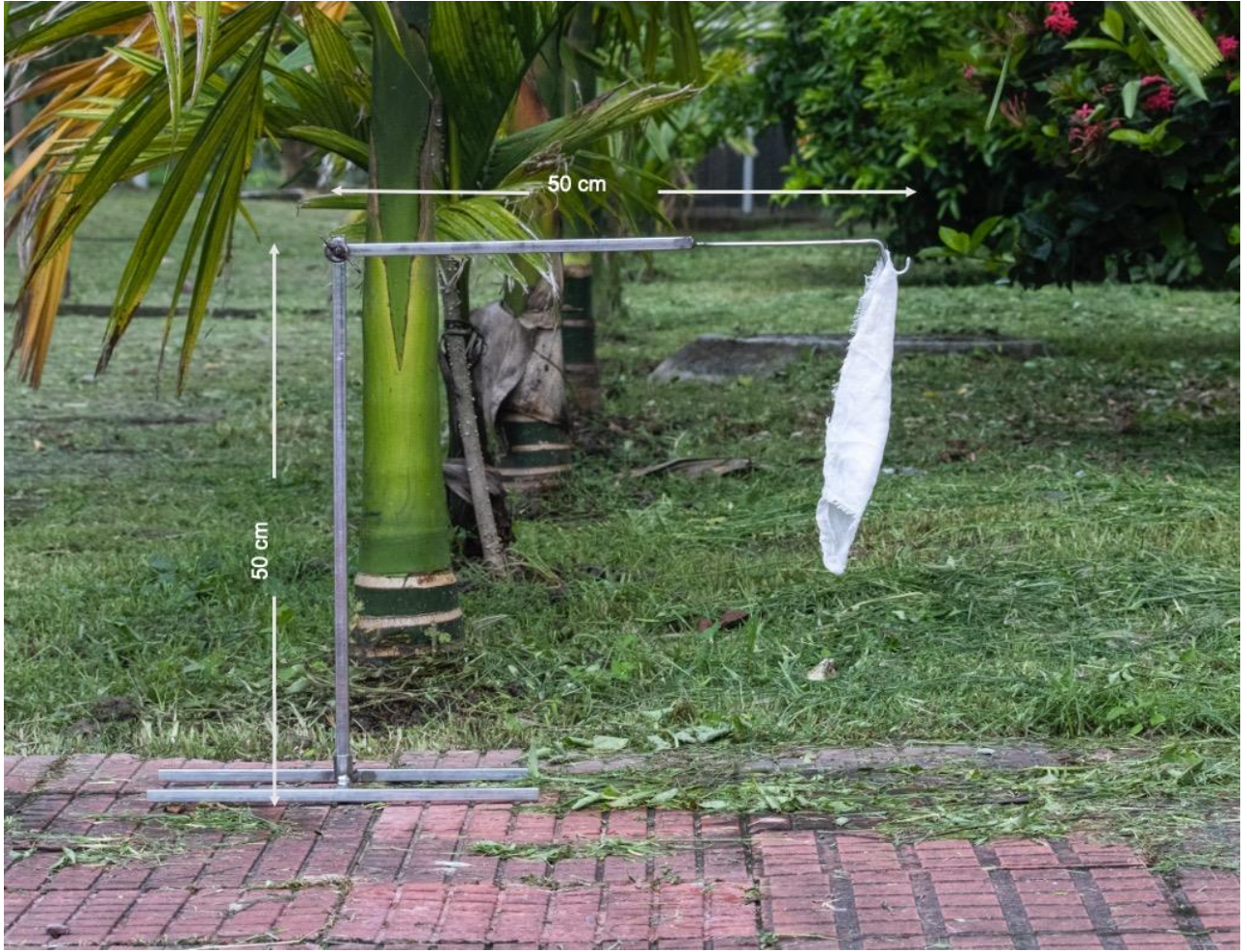

**Fig.S2: Experimental setup:** This stand measured 50 cm in height and featured a 50 cm extended arm with a hook for attachment of cloth. During the experimental trials, dog groups were presented with either a test cloth with dog scent mark or a control cloth, with a splash of water.

**Table. S2a:** The effects of sample category on the investigation time of first responder dog.

**GLM:**

**Formula:** Investigation by first responder ~ Sample category (scent mark/control)

| Parameter | Estimate | Std. Error | t value | p-value |
|---|---|---|---|---|
| (Intercept) | 1.3322 | 0.1762 | 7.562 | **< 0.001** |
| Sample category: scent mark | 0.8290 | 0.1994 | 4.157 | **< 0.001** |

**Table. S2b:** The effects of responder gender, experimental group, sample gender and experimental position in the territory on the investigation time of first responder dog.

**GLM:**

**Formula:** Investigation by first responder ~ Gender of the responder (male/female) + Experimental group + Scent mark sample gender (male/female) + Position of territory of experiment (resource areas / resting areas / territorial boundaries)

| Parameter | Estimate | Std. Error | t value | p-value |
|---|---|---|---|---|
| (Intercept) | 1.74756 | 0.34815 | 5.020 | **< 0.001** |
| Responder gender: male | -0.03064 | 0.17936 | -0.171 | > 0.05 |
| Experimental group: neighbour | 0.73223 | 0.20382 | 3.593 | **< 0.001** |
| Sample gender: male | -0.10209 | 0.29837 | -0.342 | > 0.05 |
| Position: resting | -0.21828 | 0.20392 | -1.070 | > 0.05 |
| Position: territorial border | 0.10082 | 0.24000 | 0.420 | > 0.05 |

**Table. S3a:** The effects of sample category on the average investigation time by all responder dogs.

**GLM:**

**Formula:** Average Investigation time by all responder ~ Sample category (scent mark/control)

| Parameter | Estimate | Std. Error | t value | p-value |
|---|---|---|---|---|
| (Intercept) | 1.3942 | 0.1426 | 9.779 | < 0.001 |
| Sample category: scent mark | 0.8219 | 0.1611 | 5.102 | < 0.001 |

**Table. S3b:** The effects of responder gender, experimental group, sample gender and experimental position in the territory on the average investigation time by all responder dogs.

**GLM:**

**Formula:** Average Investigation time by all responder ~ Gender of the responder (male/female) + Experimental group + Scent mark sample gender (male/female) + Position of territory of experiment (resource areas / resting areas / territorial boundaries)

| Parameter | Estimate | Std. Error | t value | p-value |
|---|---|---|---|---|
| (Intercept) | 1.8725 | 0.2737 | 6.843 | < 0.001 |
| Responder gender: male | 0.1838 | 0.1435 | 1.281 | > 0.05 |
| Experimental group: neighbour | 0.5927 | 0.1651 | 3.591 | < 0.001 |
| Sample gender: male | -0.2820 | 0.2268 | -1.243 | > 0.05 |
| Position: resting | -0.0621 | 0.1634 | -0.380 | > 0.05 |
| Position: territorial border | 0.1483 | 0.1972 | 0.752 | > 0.05 |

**Table. S4a:** The effects of sample category on territorial behaviour response scores.

**GLM:**

**Formula:** Territorial behaviour response scores ~ Sample category (scent mark/control)

| Parameter | Estimate | Std. Error | t value | p-value |
|---|---|---|---|---|
| (Intercept) | 1.2085 | 0.1300 | 9.297 | < 0.001 |
| Sample category: scent mark | 0.4124 | 0.1467 | 2.812 | < 0.01 |

**Table. S4b:** The effects of experimental group, experimental position in the territory and responder gender on territorial behaviour response scores.

**GLM:**

**Formula:** Territorial behaviour response scores ~ Experimental group (male/female) + Position of territory of experiment (resource areas / resting areas / territorial boundaries) + Gender of the responder (male/female).

| Parameter | Estimate | Std. Error | t value | p-value |
|---|---|---|---|---|
| (Intercept) | 1.56970 | 0.16470 | 9.531 | < 0.001 |
| Experimental group: same | -0.29651 | 0.17106 | -1.733 | > 0.05 |
| Position: resting | -0.17950 | 0.17067 | -1.052 | > 0.05 |
| Position: territorial border | 0.00582 | 0.20463 | 0.028 | > 0.05 |
| Responder gender: male | 0.46346 | 0.15038 | 3.082 | < 0.01 |